\documentstyle[12pt]{article}

\advance\textheight 12pt

\newcommand\bd{\begin{displaymath}}
\newcommand\ed{\end{displaymath}}
\newcommand\be{\begin{equation}}
\newcommand\ee{\end{equation}}

\begin{document}

\begin{center}
\large{{\bf Embedding initial data for black hole collisions}}
\end{center}

\vspace{20pt}
\begin{center}
by\\Joseph D. Romano and Richard H. Price \\Department of
Physics\\University of Utah\\Salt Lake City UT 84112
\end{center}

\section*{Abstract}

The visualization of curved sections of spacetime can be of considerable
conceptual value.
We discuss here the visualization of initial data for the problem of
the head-on collision of two black holes.
The problem of constructing the embedding diagram is explicitly
presented for the best studied initial data, the Misner geometry.
We present a partial solution of the embedding diagrams and discuss
issues related to completing the solution.

\bigskip
\noindent PACS number(s): 0420, 0240, 0270, 9760L

\section*{1. Introduction}

A benchmark for numerical relativity has been the computation of the
gravitational radiation waveforms generated by the collision of two
Schwarzschild throats starting from rest\cite{annetal}. This
description of the starting configuration is not complete. One can
choose initial data in many ways to fit this verbal description.
What remains to be specified can be said to correspond to the initial
mutual distortion of the holes or the long-wavelength radiation
present on the initial hypersurface. (Initial short wavelength
radiation would presumably be ``obvious.'')  When the initial
separation of the throats is sufficiently small, a single
nearly-spherical horizon surrounds both throats, and the exterior
geometry can be thought of as a perturbation of a single Schwarzschild
throat. Recent studies\cite{PP} show that this viewpoint is reasonably
successful even when the horizon is highly distorted or is even split
into two disjoint horizons. The extent to which the initial geometry
mixes its ``two throat'' and ``one throat'' nature is the key to
understanding this perturbation approach which promises to be
important in the future. One would like to get a feeling for how this
changes with changing separation of the throats, with different
choices of initial data for a given separation, etc.  For these, and
many other reasons, it would be very useful to have a direct way of
visualizing initial data. Indeed, computer visualization would seem
most appropriate for data that has been the starting point for so much
intensive numerical computation.

This would certainly seem possible in principle. The initial data
under discussion are all momentarily stationary, so---in appropriate
4-dimensional coordinates---the initial time rate of change of the
metric is zero; the initial data then consists only of the initial
3-geometry. Since that geometry is rotationally symmetric about the
symmetry axis along which the holes will move, we can take a slice
through that axis. The initial data is then fully specified by the
spatial 2-geometry on that slice. A curved 2-geometry can, at least
locally, be represented isometrically by a curved surface
in flat Euclidean 3-space. Due to the nature of the black hole initial
data one supposes such a surface to have the general shape of a pair
of trousers. Such pictures in fact are commonly drawn, but the actual
surface, to our knowledge, has never been computed.

In the present paper we discuss the generation of an embedding diagram
for the most commonly used black hole initial data, the ``Misner
data'' \cite{Misner}. On rather general grounds we show that the
generation of the embedding cannot proceed smoothly. We show
explicitly that for the Misner geometry the breakdown takes the form
of the mathematical equivalent of shock waves in the embedding
surface. This breakdown occurs, however, only quite near the crotch of
the trousers and does not stop us from computing most of the embedding
surface.  This ``partial'' embedding is sufficient for most aspects of
visualization of the initial data.  In a subsequent paper we will
report on the extent to which it is possible to overcome the obstacles
to a global embedding.

The remainder of the paper is organized as follows: Section 2
introduces the mathematical preliminaries for the general problem of
embedding a curved 2-dimensional geometry in Euclidean 3-space. Special
emphasis is given to the case where the equations are everywhere
hyperbolic since that turns out to be the case for the Misner
geometry, which is described in Sec.~3. Numerical results are given
in Sec.~4 along with a description of how the solution of the embedding
breaks down at the formation of a ``shock.'' In Sec.~5 we discuss
whether these shocks are inevitable and, if they are, how the
incomplete embedding diagrams can be of use.  As a specific example,
we show how they aid in the understanding of the range of validity of
perturbation theory.

\section*{2. The mathematics of embedding:\\
The Darboux equation}

Consider a positive-definite, 2-dimensional geometry described in terms
of local coordinates $(x,y)$ and line element
\begin{equation}
ds^2=E\,dx^2+2F\,dx\,dy+G\,dy^2\,.
\label{eq:ds^2}
\end{equation}
We would like to realize this abstract 2-geometry as a curved surface in
flat Euclidean 3-space, subject to the condition that the line element
induced on the surface by the flat Euclidean geometry agrees with that
of (\ref{eq:ds^2}).
We want, in other words, to find three functions $(U,V,W)$ such that
\begin{equation}
du^2+dv^2+dw^2=E\,dx^2+2F\,dx\,dy+G\,dy^2
\end{equation}
when $u=U(x,y)$, $v=V(x,y)$, $w=W(x,y)$ are substituted into the left
hand side of the above equation.
This requirement leads to three conditions
\begin{eqnarray}
E&=& U_{,x}^2+V_{,x}^2+W_{,x}^2\label{eq:E}\\
F&=& U_{,x}U_{,y}+V_{,x}V_{,y}+W_{,x}W_{,y}\label{eq:F}\\
G&=& U_{,y}^2+V_{,y}^2+W_{,y}^2\label{eq:G}
\end{eqnarray}
where $U_{,x}$ means partial derivative of $U$ with respect to $x$, etc.
The functions $(U,V,W)$ are called embedding funtions, and the
mapping $(x,y)\rightarrow(U,V,W)$ defines a 2-dimensional surface in
the $(u,v,w)$ Euclidean 3-space.
A set of functions $(U,V,W)$ satisfying (\ref{eq:E})-(\ref{eq:G}) is
referred to in the mathematical literature as a locally isometric
embedding.
Our goal is to find such an embedding for the 2-geometry corresponding
to the Misner initial data.

Rather than try to solve the above system
of nonlinear, first-order partial
differential equations (PDEs) for the unknowns $(U,V,W)$, we will take
a different approach originally due to Darboux.  (See, for example,
\cite{Spivak5}.)  This will lead to a single, nonlinear, second-order
PDE for the embedding function $W(x,y)$.  Although the resulting
equation is nonlinear in the second partial derivatives of $W$, it is
of a special type and can be reduced to a characteristic system of
five quasilinear, first-order PDEs.  It is this system of equations
which we then try to solve numerically.  Once we find $W$, the other
embedding functions $U$ and $V$ are determined, in terms of $W$, by
means of quadratures.  The purpose of the remainder of this section
(and the appendix at the end of the paper) is to make these statements
more precise.

Consider, instead of (\ref{eq:ds^2}), the 2-dimensional line element
\begin{equation}
du^2+dv^2=(E\,dx^2+2F\,dx\,dy+G\,dy^2\,)-dw^2\,.
\label{eq:flat1}
\end{equation}
Substituting $w=W(x,y)$ into the right hand side of the above equation,
we find
\begin{equation}
du^2+dv^2=(E-W_{,x}^2)\,dx^2+2(F-W_{,x}W_{,y})\,dx\,dy+(G-W_{,y}^2)\,dy^2\,.
\label{eq:flat2}
\end{equation}
Without loss of generality, we can assume that $W_{,x}$ and $W_{,y}$
vanish at the point about which we are trying to find the locally isometric
embedding.
(We can always perform a Euclidean transformation---i.e., a translation
and rigid rotation---to guarantee that this is actually the case.)
Since $du^2+dv^2$ is positive-definite, and hence nondegenerate, in a
neighborhood of that point, it follows that (\ref{eq:ds^2}) must be
nondegenerate for it to be be locally embeddable.
The left hand side of (\ref{eq:flat2}) is also flat, and hence has
vanishing Gaussian curvature.
(There are no additional geometric constraints.
In two dimensions, the Riemann curvature tensor is determined completely
by the scalar curvature, and the Gaussian curvature equals one-half the
scalar curvature.)
If we express the Gaussian curvature of (\ref{eq:flat2}) in terms of the
appropriate first and second partial derivatives of the components
\begin{equation}
(E-W_{,x}^2)\,,\,\,\,(F-W_{,x}W_{,y})\,,\,\,\,\,(G-W_{,y}^2)
\end{equation}
we find
\begin{equation}
{\cal A}(rt-s^2)+{\cal B}r+{\cal C}s+{\cal D}t+{\cal E}=0
\label{eq:maeqn}
\end{equation}
where ${\cal A,B,C,D,E}$ are complicated functions of $E,F,G$ and
their first and second partial derivatives, and of the first partial
derivatives $p:=W_{,x}$ and $q:=W_{,y}$ of $W$.  (See the appendix for
explicit expressions for ${\cal A,B,\cdots,E}$.)  Here $r$, $s$, $t$
are shorthand notations for the second partial derivatives $W_{,xx}$,
$W_{,xy}$, $W_{,yy}$.  All third derivatives of $W$ cancel when
calculating the Gaussian curvature of (\ref{eq:flat2}).  A
second-order PDE of the general form (\ref{eq:maeqn}), with ${\cal
A,B,\cdots,E}$ independent of $r,s,t$, is said to be of the
Monge-Amp\`{e}re type.  It differs from the general, second-order PDE
in that the only nonlinearity in $r,s,t$ occurs in the combination
$(rt-s^2)$. This special feature results in a number of important
simplifications below.  In the context of the embedding problem---with
${\cal A,B,\cdots,E}$ depending on $E,F,G$ as in the appendix---equation
(\ref{eq:maeqn}) is called the Darboux equation.

If we write the general, nonlinear, second-order PDE for the unknown
$W(x,y)$ as
\begin{equation}
Q(x,y,W,p,q,r,s,t)=0
\label{eq:PDE}
\end{equation}
then the characteristic curves $(x(\lambda),y(\lambda))$ are those
for which
\begin{equation}
Q_{,r}\left(\frac{dy}{d\lambda}\right)^2
-Q_{,s}\left(\frac{dx}{d\lambda}\right)
\left(\frac{dy}{d\lambda}\right)
+Q_{,t}\left(\frac{dx}{d\lambda}\right)^2
=0.
\label{eq:quadf}
\end{equation}
The discriminant
\begin{equation}
\Delta:=Q_{,s}^2-4Q_{,r}Q_{,t}
\label{eq:Delta}
\end{equation}
determines whether real solutions exist for
$(x(\lambda),y(\lambda))$. An equation is said to be hyperbolic
for regions in which $\Delta>0$, so that two directions exist for
characteristics at each point in the region. The equation is said to
be elliptic for $\Delta<0$, and parabolic for $\Delta=0$.

In the case of the Monge-Amp\`{e}re equation
\begin{eqnarray}
Q_{,t}&=&{\cal D}+{\cal A}r\label{eq:Qt}\\
Q_{,s}&=&{\cal C}-2{\cal A}s\label{eq:Qs}\\
Q_{,r}&=&{\cal B}+{\cal A}t\label{eq:Qr}
\end{eqnarray}
yielding
\begin{equation}
\Delta={\cal C}^2-4{\cal BD}+4{\cal A}{\cal E}\,.
\label{eq:maDelta}
\end{equation}
The above expression for $\Delta$ does not contain $r$, $s$,
or $t$.

In the case of the Darboux equation, $\Delta$ has a remarkable
geometric property: It can be written in terms of the Gaussian
curvature $K$ of the original line element in (\ref{eq:ds^2}):
\begin{equation}
\Delta=-16\,K\,(EG-F^2)^3\,(n^3)^2\,.
\label{eq:DarbouxDelta}
\end{equation}
(Here $n^3$ is the $w$-component of the unit normal to the embedded
surface. See the appendix for details.)
For a positive-definite 2-geometry, $EG-F^2>0$ and the sign of
$\Delta$ is determined by the Gaussian curvature $K$.
For $K>0$ the Darboux equation is elliptic, so there are no
characteristic curves.
This means that features on one part of the embedding surface
influence all other parts of the embedding.
As we will show in the next section and in the appendix,
the 2-geometry for the Misner initial data has everywhere-negative
Gaussian curvature.
In this case, the Darboux equation is hyperbolic, and features,
such as the choice of boundary conditions for the embedding,
propagate along characteristics and remain fairly localized.

A geometric meaning for the characteristics of the Darboux equation
emerges if we use (\ref{eq:Qt})-(\ref{eq:Qr}) and the expressions
for ${\cal A},{\cal B},\cdots,{\cal E}$ given in the appendix to
rewrite (\ref{eq:quadf}) as
\begin{eqnarray}
\lefteqn{Q_{,r}\left(\frac{dy}{d\lambda}\right)^2
-Q_{,s}\left(\frac{dx}{d\lambda}\right)
\left(\frac{dy}{d\lambda}\right)
+Q_{,t}\left(\frac{dx}{d\lambda}\right)^2}
\nonumber\\
& &=-4\,n^3(EG-F^2)\left[K_{11}\left(\frac{dx}{d\lambda}\right)^2
+2K_{12}\left(\frac{dx}{d\lambda}\right)
\left(\frac{dy}{d\lambda}\right)
+K_{22}\left(\frac{dy}{d\lambda}\right)^2
\right].\nonumber
\end{eqnarray}
Here $K_{11},K_{12},K_{22}$ are the $(x,y)$-components of the extrinsic
curvature tensor ${\bf K}$ for the embedding (see the appendix).
The characteristic directions for the 2-geometry of (\ref{eq:ds^2})
are thus seen to be the zero vectors (i.e., asymptotic directions) of
${\bf K}$; they are the vectors $\vec{v}$ for which
$\vec{v}\cdot{\bf K}\cdot\vec{v}=0$.  Clearly, for $K<0$ there are two
principal curvatures of opposite sign. The principal directions are
orthogonal in the geometry of (\ref{eq:ds^2}) and it is easy to see
that the characteristic directions must be symmetrically arranged with
respect to the principal directions.  That is, a principal direction
must bisect the angle between a pair of characteristics.

Since we are guaranteed that there exist two characteristics through any
point, we can use the characteristics themselves as coordinates.  We
label one family of characteristics (i.e., one family of solutions of
(\ref{eq:quadf})) with $\alpha$ and the other family with $\beta$.
The Darboux equation can then be reformulated in terms of five unknown
functions $(x,y,W,p,q)$ of the variables $(\alpha,\beta)$.
It is shown in the appendix that these are determined by the five
quasilinear, first-order PDEs
\begin{eqnarray}
{1\over 2}({\cal C}-\delta)\,x_{,\alpha}-{\cal B}\,y_{,\alpha}
-{\cal A}\,q_{,\alpha}&=&0
\label{eq:cs1}\\
{1\over 2}({\cal C}+\delta)\,x_{,\beta}-{\cal B}\,y_{,\beta}
-{\cal A}\,q_{,\beta}&=&0
\label{eq:cs2}\\
{1\over 2}({\cal C}+\delta)\,y_{,\alpha}-{\cal D}\,x_{,\alpha}
-{\cal A}\,p_{,\alpha}&=&0
\label{eq:cs3}\\
{1\over 2}({\cal C}-\delta)\,y_{,\beta}-{\cal D}\,x_{,\beta}
-{\cal A}\,p_{,\beta}&=&0
\label{eq:cs4}\\
W_{,\alpha}-p\,x_{,\alpha}-q\,y_{,\alpha}&=&0\label{eq:cs5}
\end{eqnarray}
where $\delta:=\sqrt{\Delta}$. (It should be noted that the general,
nonlinear, second-order hyperbolic PDE, in characteristic form,
requires eight equations for the eight unknowns
$(x,y,W,p,q,r,s,t)$. It is the special nature of the Monge-Amp\`{e}re
equation that results in a system of only five equations\cite{C&H2}.)

Of the three embedding functions $(U,V,W)$, the Darboux approach
singles out one, $W$. It turns out, as shown in the appendix,
that the remaining steps to complete the embedding are fairly
straightforward.
Once the solutions to (\ref{eq:cs1})-(\ref{eq:cs5}) are found,
the remaining embedding functions $U,V$ are determined in terms of
$W$ by means of quadratures\cite{H&W}.

To close this section, we point out that the method of Darboux described
above is only one of many approaches to solving the embedding problem.
There is an extensive Russian mathematical literature on
locally isometric embeddings for negative Gaussian curvature 2-geometries
which describes the method of Riemann invariants.
Interested readers should see the review article by Poznyak and
Shikin\cite{P&S} for more information.
Also, a paper by Bernstein\cite{Bernstein} describes an iterative
numerical scheme for computing isometric embeddings.

\section*{3. The Misner geometry}

The  black hole initial data used for numerical relativity
studies\cite{annetal} is the Misner 3-geometry\cite{Misner} given by
\begin{equation}
ds^{2}_{\rm Misner}=a^{2}\,\varphi^{4}_{\rm Misner}
\left[\,d\mu^{2}+d\eta^{2} +\sin^2{\eta}\,d\phi^{2}\,\right]
\label{eq:ds^2misner}
\end{equation}
where
\begin{equation}
\varphi_{\rm Misner}=
\sum_{n=-\infty}^\infty\frac{1}{\sqrt{\cosh(\mu+2n\mu_0)-\cos\eta}}\,.
\label{eq:phimisner}
\end{equation}
The geometry has a total ADM mass
\begin{equation}
M_{\rm tot}=4a\,\sum_{n=1}^\infty\,{1\over\sinh(n\mu_0)}
\label{eq:Mtot}
\end{equation}
and describes two throats located near $\mu=\pm\mu_0$.
A measure of the separation of the throats is $L$, the proper distance
from $\mu=-\mu_0$ to $\mu=+\mu_0$, which can be shown to be
\begin{equation}
L=2a\left(1+2\mu_0\sum_{n=1}^\infty\frac{n}{\sinh(n\mu_0)}\right)\,.
\end{equation}
The constant $a$ is a scaling factor that sets the size of both
$M_{\rm tot}$ and $L$, but not the ratio $L/M_{\rm tot}$, and does not
affect the ``shape'' of the geometry. The only parameter affecting the
shape is $\mu_0$, which is an increasing function of $L/M_{\rm tot}$.
For $\mu_0\ll1$ there is a single, nearly spherical, initial horizon;
for $\mu_0\gg1$ the geometry represents two widely separated throats,
each with an initial horizon.
The transition from a single horizon to a split horizon with two
disjoint segments occurs at $\mu_0\approx1.8$.

The Misner 3-geometry is rotationally symmetric---i.e., it is
independent of the the azimuthal angle $\phi$.
We therefore lose no geometric
information by taking a $\phi=const$ slice of (\ref{eq:ds^2misner}),
thereby arriving at a 2-geometry
\begin{equation}
ds^2=a^2\,\varphi_{\rm Misner}^4\left[\,d\mu^2+d\eta^2\,\right]
\label{eq:misner2}
\end{equation}
and the possibility of an embedding in Euclidean 3-space.

A numerical study of the Gaussian curvature $K$ of this 2-geometry
(\ref{eq:misner2}) shows that the Gaussian curvature is everywhere-negative.
We have also been able to write the Gaussian curvature in a form in which
it is manifest that it is everywhere-negative:
\begin{eqnarray}
\lefteqn{K=-{1\over 2}\,a^{-2}\,\varphi_{\rm Misner}^{-6}}\nonumber\\
& &\sum_{m=-\infty}^\infty\sum_{n=-\infty}^\infty
{\cosh\left(2\mu_0(m-n)\right)-1\over
\left[\cosh(\mu+2m\mu_0)-\cos\eta\right]^{3/2}
\left[\cosh(\mu+2n\mu_0)-\cos\eta\right]^{3/2}}\,.\nonumber
\end{eqnarray}
(See the appendix for details.)

For our analysis,
it is convenient to transform from the $(\mu,\eta)$ coordinates to
coordinates better suited to the description of the geometry at large
distances from the holes.
To do this we transform from $(\mu,\eta)$ to $(\theta,R)$, as if from
bispherical to polar coordinates, according to
\begin{eqnarray}
\theta&=&\arctan\left({\sin\eta\over\sinh\mu}\right)\\
R&=&{a\over\cosh\mu-\cos\eta}\sqrt{\sinh^2\mu+\sin^2\eta}\,.
\end{eqnarray}
The inverse transformation formulae are
\begin{eqnarray}
\mu&=&\pm{\rm arccosh}\left(\frac{R^2+a^2}{\sqrt{(R^2+a^2)^2-
(2aR\cos\theta)^2}}\right)\\
\eta&=&\pm{\rm arccos}\left(\frac{R^2-a^2}{\sqrt{(R^2-a^2)^2+
(2aR\sin\theta)^2}}\right)\,.
\end{eqnarray}
The signs are determined by requiring that $\mu>0$ for
$-\pi/2<\theta<+\pi/2$ and $\eta>0$ for $0<\theta<\pi$.
In terms of the $(\theta,R)$ coordinates, the 2-geometry
(\ref{eq:misner2}) takes the form
\begin{equation}
ds^2=\Phi^4(\theta,R)\left(R^2\,d\theta^2+dR^2\right)
\label{eq:ds^2conformal}
\end{equation}
with
\begin{equation}
\Phi=1+\frac{a}{R}\sum_{n\not=0}
\frac{1}{\,|\sinh(n\mu_0)|\sqrt{1+(2a/R)\coth(n\mu_0)\cos\theta
+(a^2/R^2)\coth^2(n\mu_0)}}\,.
\label{eq:calF}
\end{equation}
If we identify $(\theta,R)$ with $(x,y)$, then the coefficients
$E,F,G$ of the line element (\ref{eq:ds^2}), which we take as the
starting point for the embedding problem, are given by
\begin{eqnarray}
E&=&R^2\,\Phi^4(\theta,R)\\
F&=&0\\
G&=&\Phi^4(\theta,R)\,.
\end{eqnarray}

Since $K<0$ for the Misner 2-geometry, the corresponding Darboux equation
is hyperbolic and appropriate Cauchy data must be specified.
We choose to specify ``initial'' data at large constant $R$.
Here the 2-geometry (\ref{eq:ds^2conformal}) is approximately that
of a single, central Schwarzschild throat, and $R$ plays the role of the
Schwarzschild isotropic radius.
$R$ is related to the Schwarzschild curvature radius $r_{\rm curv}$
via
\begin{equation}
R={1\over 2}\,r_{\rm curv}\left(1-M_{\rm tot}/r_{\rm curv}
+\sqrt{1-2M_{\rm tot}/r_{\rm curv}}\right)
\end{equation}
where $M_{\rm tot}$ is the total ADM mass (\ref{eq:Mtot}) of the single,
central Schwarzschild throat.
Equivalently,
\begin{equation}
r_{\rm curv}=R\,\left(1+M_{\rm tot}/2R\right)^2\,.
\end{equation}
For a single Schwarzschild throat of mass $M_{\rm tot}$, the embedding
is given by
\begin{equation}
W=2M_{\rm tot}\left(\sqrt{2R/M_{\rm tot}}-\sqrt{M_{\rm tot}/2R}\right)\,.
\label{eq:schwemb}
\end{equation}
The initial values of $W,p,q,r,s,t$ on the $R=const$ surface are
found from (\ref{eq:schwemb}) and its derivatives with respect to
$\theta$ and $R$.

Our numerical approach is a finite-difference solution to the
characteristic equations in (\ref{eq:cs1})-(\ref{eq:cs5}).
On the $R=const$ initial value surface, grid points are chosen to be
equally spaced in $\theta$.
A pair of $(\alpha,\beta)$ characteristics is then started from
each initial grid point.
Equations (\ref{eq:cs1})-(\ref{eq:cs5}) are used to propagate
$(x,y,W,p,q)$ forward along the characteristics. The two axes of
bilateral symmetry of the Misner geometry are used to reduce the size
of the numerical grid by a factor of four.  The infinite sums were
approximated by finite sums from $-N_{\rm sum}$ to $N_{\rm sum}$, with
$N_{\rm sum}$ large enough so that the omitted terms were negligible.

\section*{4. Results: Characteristics and partial embeddings for the
Misner geometry}

Equations (\ref{eq:cs1})-(\ref{eq:cs5}) were solved numerically, and
from the results, solutions were constructed for
$U(\alpha,\beta), V(\alpha,\beta)$.
With these solutions and the solution for $W(\alpha,\beta)$, the two
families of characteristics in the $(u,v,w)$ space are known.
The results are shown in Fig.~1 in the three cases $\mu_0=1,2,3$ which
correspond to $L/M_{\rm tot}=1.92,3.88,7.92$. In these figures and
in all those below, the value of the scaling constant $a$ was chosen,
for each value of $\mu_0$, so that $M_{\rm tot}=1$. The $(u,v)$
coordinates therefore measure distances in units of $M_{\rm tot}$, so
that in the figures we are always visually comparing spacetimes
with the same total mass, but with throats at different separation.

An immediately apparent, and crucial, feature of the figures is that
the characteristic net does not cover the complete $(u,v)$ interior to the
$R=const$ initial value surface. The numerical solution inevitably breaks
down at the boundary of an oval region that includes the throats. This
is not a numerical artifact; extensive numerical testing confirmed
that the location of the breakdown was reasonably independent of
numerical step size.

A breakdown is, in fact, mathematically inevitable. This can most
easily be seen by imagining the characteristics at the midpoint of
the $(u,v)$ plot. Since the underlying
2-geometry is symmetric about the rotation axis, the embedding diagram
must be reflection symmetric about the horizontal line through the
midpoint.  The symmetry axis, then, must be a principal direction of the
extrinsic curvature. The zero directions
of ${\bf K}$, and hence the directions of the characteristics, must be
symmetrically arranged about the principal direction.
The nature of the characteristics must therefore be that shown in Fig.~2.
But what direction along these
characteristics is the ``forward'' direction of propagation of
information from the boundary?  A choice of direction along these
characteristic segments violates the symmetry of the boundary.
The direction of forward propagation must be ill-defined at the midpoint.
An extension of this argument shows that this must be a problem
not only at the midpoint but also, at least, on the symmetry axis
joining the throats.
The inescapable conclusion is that the characteristic net emanating from
the $R=const$ boundary cannot smoothly and continuously cover the entire
region inside the boundary and outside the throats.

This conclusion, in fact, does not depend on the symmetry of the
Misner geometry, but only on its general topological character. We can
shrink the $R=const$ boundary, and the throats, so that the embedding
region becomes a 2-sphere with three singular points, representing the
boundary and each of the throats.  Let us suppose that we can put a
single family of characteristics on this 2-sphere, starting from the
boundary and ending only at the throats. The tangent vectors to the
characteristics (pointing in the direction of propagation of the solution)
constitute a vector field having no singularities or zeroes except at
the boundary and at the throats.  At the boundary, the tangent vectors
all point outward, while at the throats, they all point inward (if the
specification of additional boundary conditions is to be advoided).
The index of the vector field\cite{Hsiung} would then be +1 at each of
these points, and the total index of the vector field would equal 3.
But the Poincar\'e index theorem\cite{Hsiung} requires that the index of
a vector field on a manifold equal the Euler characteristic $\chi$ of
the manifold. For a 2-sphere, the Euler characteristic $\chi=2$, so the
tangent field to the family of characteristics is impossible.  A
simple generalization of this argument shows that problems must
develop in the embedding of negative Gaussian curvature initial data
for any number of holes other than unity.

It remains to determine the precise nature of the inescapable
breakdown in the propagation of the characteristics. It cannot be due to a
geometric singularity. The {\em intrinsic} geometry of the embedded surface
is fixed, of course, by the requirement that it be isometric
to the nonsingular Misner 2-geometry.  We are carefully studying the
{\em extrinsic} geometry of the embedded surface that we generate
numerically, and we have found no evidence, so far, that any singularity
is developing as the characteristics approach the point of breakdown.

The actual nature of the breakdown is suggested by our argument above
about the incompatible directions of characteristics.  The breakdown
occurs when characteristics of the same family cross each other,
as in the classic example of shocks in gas flow. At such
crossings there are three characteristic directions (two from the
crossed characteristics, and one from the characteristic of the other
family) and the propagation of the solution cannot proceed.  Figure 3
shows a detail of the numerical results for the propagation of
characteristics in the extreme case $\mu_0=6$. To the left of the crossing
at $u\approx 6.6, v\approx 8.2$ the results are valid; to the right,
the results are meaningless.

Although the development of these ``embedding shocks'' (i.e., characteristic
crossings) block the computation of the embedding in an inner region
of the geometry, we can still compute the outer region and obtain
most of the visually useful information. Figure 4 shows the outline of the
embedded surface (a view along the $v$ axis) for the cases $\mu_0=1$ and
$\mu_0=3$. A clear geometric distinction is apparent.  For the $\mu_0=3$
case, the structure close in (in the region $-5<u<5$) is that of two
throats.  On a larger
scale the geometry takes on the character of a single Schwarzschild
throat. The shape of the $\mu_0=1$ case is quite different. Though it
has the same ``trousers'' topology as that of $\mu_0=3$, the steep
sides of the embedding near $u=\pm2$ mean that there is no region in which
the individual throats look like more-or-less isolated holes. A
3-dimensional plot of the $\mu_0=3$ geometry is presented in Fig.~5.

\section*{5. Discussion}

The development of ``embedding shocks" is the most interesting feature of
the problem of embedding the everywhere-negative Gaussian curvature
Misner 2-geometry. The primary question to be asked about them is
whether it is inevitable that a ``forbidden region" develop towards the crotch
of the embedding, cut off from the outer region by the embedding shock. The
topological arguments of the previous section clearly establish that {\em
something} must go wrong with the characteristics, but the failure could be
confined to the symmetry axis joining the throats, at which the
characteristics meet with incompatible directions of forward propagation.
In effect, the ``forbidden region" could be degenerate, with zero area.
This is not, of course,
what our numerical results show, but the location of the embedding shock
depends on the initial data. A correct small change in the initial data might
shift the location of the embedding shock inward, reducing the area of the
forbidden region; the precisely correct initial data might even reduce the
forbidden region to a line. We have numerically studied the effect on the
shock location of the precise form of the initial data.
The results strongly suggest that no improvement of the Cauchy data can
significantly reduce the size of the forbidden region. [There is an
additional point that must be added in connection with this. Part of the
problem in the oval-shaped breakdown may be a technical difficulty that
develops when the generated surface becomes vertical. This would be most
expected near the $v=0$ plane at the outer edge of the throats.
This problem, however, is causally
disconnected from the breakdown near the $u=0$ plane, so the breakdown
cannot wholly be a simple technical flaw in the approach. We are
also studying this point, and developing a numerical technique for avoiding
this problem, which may be inherent to the Darboux approach.]

Suppose that somehow or another we {\it were} able to reduce the
size of the forbidden region to that of zero area.
What would be the implication of such a degenerate embedding shock?
It would mean that the embedding diagram
is uniquely determined by the initial data on the (approximately) circular
outer boundary. That information would propagate inward unambiguously
creating the embedding surface in its wake, until it hits the central
degenerate shock. Whether the surface generated from one side and that
generated from the other meet smoothly at this line is not guaranteed.
The degenerate embedding shock would then be a sort of zero measure
failure of the local isometric embedding to be global. This leads us to point
out that theorems on the existence of global embeddings are notoriously
nonexistent. There is no {\em a priori} reason to suspect that a global
embedding of the Misner geometry either does or does not exist.

If a global embedding does not exist, then an attempt to start at an outer
circular boundary and propagate inward is doomed to incompleteness. We
could, of course, start with inner data, and propagate the embedding diagram
outward. But to discuss why we choose not to do this,
it is useful at this point to
review what we are trying to accomplish with the embedding. To help in the
visualization of the physics, the embedding diagram must have the character
of a single throat at large radii, and in some sense must represent the idea
of two holes at smaller radii. If a global embedding is impossible, then a
diagram started from the central region would be guaranteed, at large radii,
to look nothing like the embedding diagram of a single Schwarzschild throat.
While there might be some uses to a diagram which represents the inner
regions of the Misner geometry, at least for the visualization of initial
data for black hole collisions, it would seem much more important to have
the desired features at large radii, than in the inner region.

In the discussion above, we have assumed that the forbidden region
surrounded by the embedding shock can be reduced to a line, but this may
very well not be the case.
What if the embedding shocks are unavoidable features that separate the
outer region from a nondegenerate inner region?
The mathematical parallel between
such embedding shocks and the more familiar shocks in gas flow are
revealing. In the simplest gas shocks, called ``kinematic shocks,'' the
equation of continuity, along with a relation between propagation
velocity and density, leads to a crossing of
characteristics\cite{whitham}. One then argues that the mathematical
description is only an approximation to the true physics, and that the
approximation breaks down at the shock formation. In principle, one
can turn to a more complete description of the physics, such as the
Navier-Stokes equation, to arrive at a mathematical formulation without
singularities. The simpler approximate mathematics will be adequate on
both sides of the shock; the more complete mathematical description
will tell us how to ``cross'' it. That is, how to make the physically
appropriate match of the two approximate solutions across the shock.

This sort of strategy is inapplicable to embedding shocks. The Darboux
equation is not an approximation; there is not a more complete geometric
description that can tell us how to cross the shocks ``correctly.''
But there is another view of shock-crossing conditions that may be of use.
Without reference to a more complete description of the
physics, the crossing conditions for gas shocks can be inferred from
conservations laws. In the simple case of kinematic shocks, the
crossing conditions follow when mass-conservation at the shock is
invoked. The resulting solution (without reference to more detailed
physics) is then understood to be a ``weak'' solution---i.e., one which
contains singularities, but for which the differential equation is
satisfied when it is integrated over any region, including the region of
the shock front\cite{whitham}.

What then is the geometric analogy in the case of embeddings?
This question is presently being investigated.
For now we can only point to some interesting speculations.
We note that for a sufficiently smooth surface, with extrinsic
curvature everywhere defined and continuous, there can be no shocks,
whether crossed correctly or not. It is interesting, however, to consider an
embedding function $W(u,v)$ in which there are special curves
along which  the second derivatives of $W$ are bounded but discontinuous.
Such a surface would be smooth (i.e., it would have no creases) but
along the special curves, the extrinsic curvature would be discontinuous.
At these special curves the directions of
the characteristics (which are constrained by the extrinsic curvature) would
be multiple-valued. One might imagine that such a surface would be a local
isometry to the Misner geometry except along the special curves, and it is
tempting to think that this, or some related geometric phenomenon,
would represent a weak solution to the embedding problem and the analogy of
gas shocks.

For the present, we point out that the embedding shocks do not preclude
a partial solution of the embedding problem, which is
potentially useful for visual and physical insight. An example is the
question of the applicability of perturbation theory to black hole
collisions.

If the black holes are initially close together (i.e., if $\mu_0$ is
sufficiently small) the initial geometry {\em outside the horizon} is
nearly spherical. The highly nonspherical geometry inside the horizon
does not influence the time evolution of the spacetime ouside, so the
evolution, and the generation of outgoing gravitational radiation, can
be treated as a problem in perturbations of the spherical geometry of
a single Schwarzschild throat. An important question is: How small must
$\mu_0$ be for perturbation theory to be valid? As $\mu_0$ decreases
below $\sim1.8$ the topology of the horizon on the initial Cauchy
hypersurface changes from that of two disjoint spheres (one around each
throat) to that of a single 2-sphere surrounding both throats. Only for
$\mu_0$ significantly less than 1.8 will the horizon be reasonably
spherical. Yet we know, from comparison with supercomputer evolution
of the fully nonlinear equations\cite{annetal}, that perturbation
theory works rather well for values of $\mu_0$ even somewhat larger
than 1.8. The explanation would seem to lie in the the fact that
nonsphericity just ouside the horizon disappears down the horizon. In
the theory of perturbations of the Schwarzschild
geometry\cite{sunprice2}, outgoing radiation appears to be generated
at the peak of the ``curvature potential,'' located at $r_{\rm
curv}\approx3M_{\rm tot}$. This radius, roughly, separates the regions
in which waves go outward to infinity and go inward towards the
hole. A reasonable criterion for applicability of perturbation theory
would then seem to lie in the answer to the question: How nonspherical
is the geometry near $r_{\rm curv}\approx3M_{\rm tot}$?

The difficulty in answering this question is that it requires a surface
whose sphericity can be evaluated. The horizon is a natural choice but,
of course, cannot be located at the peak of the curvature potential.
In general, other choices are not gauge invariant, being dependent on
coordinate choices.
An exception is the coordinate-independent shape of the embedding diagram.
We can then ask: How nonspherical is the embedding diagram near
$r_{\rm curv}\approx3M_{\rm tot}$?
More specifically, we can evaluate the shape of the constant $W$ contours
corresponding to $r_{\rm curv}\approx3M_{\rm tot}$ and see how they deviate
from circles.

In Fig.~6 we present such contours for $\mu_0=1.5,2,2.5$. The error
that perturbation theory makes in estimating radiation energy in these
three cases is 0, a factor of 2, and a factor of 10, respectively.
For each value of $\mu_0$, several contours are shown with the
properties summarized in Table 1 below. For those contours with
$r:=\sqrt{u^2+v^2}\approx r_{\rm curv}\approx3M_{\rm tot}$ there seems
to be a clear correlation between the success of perturabtion theory and
the spherical (that is, circular) shape of the contours.

\begin{table}[h]
\begin{tabular}{c|l|l|l|l|l|l|l|l}\hline\hline
\multicolumn{3}{c|}{$\mu_0=1.5$}&
\multicolumn{3}{c|}{$\mu_0=2$}&
\multicolumn{3}{c}{$\mu_0=2.5$}
\\ \hline
  $r_{\rm max}$ &	$r_{\rm min}$   & $\Delta r/r_{\rm max}\ ^a$ &
  $r_{\rm max}$ &	$r_{\rm min}$   & $\Delta r/r_{\rm max}$ &
  $r_{\rm max}$ &	$r_{\rm min}$   & $\Delta r/r_{\rm max}$
\\ \hline
\hspace*{10pt}2.50\hspace*{10pt}&2.36&.056&2.77&2.38&.141&3.05&2.45&.197$^b$
\\ \hline
\hspace*{10pt}2.94\hspace*{10pt}&2.86\hspace*{10pt}&.027&3.12&2.87&.080&
3.56&2.91&.183
\\ \hline
\hspace*{10pt}3.60\hspace*{10pt}&3.56&.011&3.72&3.57&.040&4.02&3.61&.102
\\ \hline
\hspace*{10pt}4.48\hspace*{10pt}&4.45&.007&4.56&4.47&.020&4.75&4.52&.048
\\ \hline\hline
\multicolumn{9}{r}{$\ ^a\ \Delta r\equiv r_{\rm max}-r_{\rm min}$
\hspace*{24pt}$\ ^b$incomplete curve}
\end{tabular}
\caption{Noncircularity of constant $W$ contours}\label{table1}
\end{table}

\section*{Acknowledgments}

We thank Andrejs Treibergs for useful suggestions and extensive
discussions of mathematical issues in connection with embeddings.
The argument we gave in Sec.~4 involving the Poincar\'{e} index
theorem arose during one of our discussions.
We also thank Ed Seidel of NCSA for valuable suggestions, and
David Bernstein of the University of New England, Australia, for
discussions of his work on the embedding problem.
JDR thanks Carsten Gundlach for help with the numerical code.
The work reported here was supported by grant NSF-PHY-92-07225,
and by research funds of the University of Utah.

\section*{Appendix}

\subsection*{A.1 Coefficients of the Darboux equation}

The coefficients ${\cal A,B,\cdots,E}$ of the Darboux equation are
\begin{eqnarray}
{\cal A}&=&-4(EG-F^2)\label{eq:A}\\
{\cal B}&=&+2p(2GF_{,y}-GG_{,x}-FG_{,y})+2q(-2FF_{,y}+FG_{,x}+EG_{,y})
\label{eq:B}\\
{\cal C}&=&+4p(FG_{,x}-GE_{,y})+4q(FE_{,y}-EG_{,x})\label{eq:C}\\
{\cal D}&=&-2p(2FF_{,x}-GE_{,x}-FE_{,y})-2q(-2EF_{,x}+FE_{,x}+EE_{,y})
\label{eq:D}\\
{\cal E}&=&+(E-p^2)(E_{,y}G_{,y}-2F_{,x}G_{,y}+G_{,x}^2)\nonumber\\
 & &+(F-pq)(E_{,x}G_{,y}-E_{,y}G_{,x}-2E_{,y}F_{,y}-2F_{,x}G_{,x}
+4F_{,x}F_{,y})\nonumber\\
 & &+(G-q^2)(E_{,x}G_{,x}-2E_{,x}F_{,y}+E_{,y}^2)\nonumber\\
 & &+2((E-p^2)(G-q^2)-(F-pq)^2)(2F_{,xy}-E_{,yy}-G_{,xx})\,.
\label{eq:EE}
\end{eqnarray}
Here $W(x,y)$ is one of the three embedding functions $(U,V,W)$, and
$p$ and $q$ are shorthand notations for the partial derivatives
$W_{,x}$ and $W_{,y}$.
The functions $E,F,G$ are the components of the 2-geometry that is
to be embedded.

\subsection*{A.2 Gaussian curvature}

The Gaussian curvature $K$ of a 2-geometry having components $E,F,G$
is given by
\begin{eqnarray}
\lefteqn{K=\left[2(EG-F^2)(-E_{,yy}+2F_{,xy}-G_{,xx})\right.}\nonumber\\
& &\,\,\,\,\,+E(G_{,x}^2-2F_{,x}G_{,y}+E_{,y}G_{,y})\nonumber\\
& &\,\,\,\,\,+F(E_{,x}G_{,y}-E_{,y}G_{,x}+4F_{,x}F_{,y}-2E_{,y}F_{,y}
-2F_{,x}G_{,x})\nonumber\\
& &\,\,\,\,\,\left.+G(E_{,y}^2-2E_{,x}F_{,y}+E_{,x}G_{,x})]/4(EG-F^2)^2
\right]\,.\label{eq:Kapp}
\end{eqnarray}
It is an intrinsic property of the 2-geometry, independent of any embedding.

\subsection*{A.3 Unit normal}

The unit normal $\vec n$ to a surface depends explicitly on the embedding
functions $\vec f:=(U,V,W)$.
Up to a normalization factor, it is given by the usual Euclidean 3-space
vector product of $\vec f_{,x}$ and $\vec f_{,y}$.
When adjusted to be of unit length and reexpressed using
(\ref{eq:E})-(\ref{eq:G}), the unit normal can be written as
\begin{equation}
\vec n={\vec f_{,x}\times \vec f_{,y}\over\sqrt{EG-F^2}}\,.
\label{eq:unitnormal}
\end{equation}
The $(u,v,w)$-components of $\vec n$ are
\begin{eqnarray}
n^1={(V_{,x}W_{,y}-V_{,y}W_{,x})\over\sqrt{EG-F^2}}\\
n^2={(W_{,x}U_{,y}-W_{,y}U_{,x})\over\sqrt{EG-F^2}}\\
n^3={(U_{,x}V_{,y}-U_{,y}V_{,x})\over\sqrt{EG-F^2}}\,.
\end{eqnarray}

\subsection*{A.4 Extrinsic Curvature}

The extrinsic curvature ${\bf K}$ is a tensor on the 2-geometry that
describes the shape of the embedded surface and depends on the
emebdding functions $\vec f=(U,V,W)$.  In terms of $\vec f$ and the
unit normal $\vec n$, its $(x,y)$-components can be written as
\begin{eqnarray}
K_{11}&=&\vec f_{,xx}\cdot\vec n\label{eq:L}\\
K_{12}&=&\vec f_{,xy}\cdot\vec n\label{eq:M}\\
K_{22}&=&\vec f_{,yy}\cdot\vec n\label{eq:N}\,.
\end{eqnarray}
Here $\cdot$ denotes the usual Euclidean 3-space dot product of two
vectors.

\subsection*{A.5 The hyperbolic Monge-Amp\`{e}re equation as a
characteristic system}

We outline here how the hyperbolic Monge-Amp\`{e}re equation can be
written as a characteristic system of quasilinear, first-order PDEs for
the unknowns $(x,y,W,p,q)$.  More details can be found in the text by
Courant and Hilbert\cite{C&H2}.

We start with the Monge-Amp\`{e}re equation
\begin{equation}
Q:={\cal A}(rt-s^2)+{\cal B}r+{\cal C}s+Dt+{\cal E}=0\,.
\label{eq:ma}
\end{equation}
This is a nonlinear, second-order PDE for a single unknown function
$W(x,y)$.
The only requirement on the coefficients ${\cal A,B,\cdots,E}$ is that they
be independent of the second partial derivatives $r:=W_{,xx}$,
$s:=W_{,xy}$, $t:=W_{,yy}$.
They can depend nonlinearly on $x,y$, and $W$, and on the first partial
derivatives $p:=W_{,x}$ and $q:=W_{,y}$.

Assume that the equation is hyperbolic, so that
\begin{equation}
\Delta:=Q_{,s}^2-4Q_{,r}Q_{,t}>0\,.
\end{equation}
In terms of the coefficients ${\cal A,B,\cdots,E}$, we have
\begin{eqnarray}
\Delta&=&({\cal C}-2{\cal A}s)^2-4({\cal B}+{\cal A}t)({\cal D}+{\cal A}r)
\nonumber\\
&=&{\cal C}^2-4{\cal BD}+4{\cal A}\,{\cal E}\,.
\end{eqnarray}
Here we used (\ref{eq:ma}) to eliminate $r,s,t$ to obtain the second
equality.
The first equality is actually an important ``identity'' that we will
use later on in our analysis.
We choose to write this identity in the form
\begin{equation}
{{\cal C}-2{\cal A}s+\delta\over 2({\cal B}+{\cal A}t)}=
{2({\cal D}+{\cal A}r)\over {\cal C}-2{\cal A}s-\delta}
\label{eq:identity}
\end{equation}
where
\begin{equation}
\delta:=\sqrt{\Delta}=\sqrt{{\cal C}^2-4{\cal B}{\cal D}
+4{\cal A}\,{\cal E}}\,.
\end{equation}
Since equation (\ref{eq:ma}) was assumed to be hyperbolic, $\delta$ is
real and is taken to be positive.

We next write down the solutions to the characteristic equation
\begin{equation}
Q_{,r}\,\left({dy\over dx}\right)^2-Q_{,s}\,\left({dy\over dx}\right)
+Q_{,t}=0\,.
\end{equation}
In terms of the coefficients ${\cal A,B,\cdots, E}$, the solutions are
\begin{equation}
\left({dy\over dx}\right){}_{\pm}={{\cal C}-2{\cal A}s\pm\delta\over
2({\cal B}+{\cal A}t)}\,.
\label{eq:dydx1}
\end{equation}
Since $\delta>0$, there exist two distinct characteristic directions
at each point.
We can integrate along these directions to obtain the characteristic
curves, which we label by two parameters $\alpha$ and $\beta$.
If we choose $\alpha$ and $\beta$ to be constant along the curves
having the directions $(dy/dx)_+$ and $(dy/dx)_-$, respectively,
then
\begin{equation}
\left({dy\over dx}\right){}_+={y_{,\beta}\over x_{,\beta}}
\quad{\rm and}\quad
\left({dy\over dx}\right){}_-={y_{,\alpha}\over x_{,\alpha}}\,.
\label{eq:dydx2}
\end{equation}
Here $x$ and $y$ are to be thought of as unknown functions of the
characteristic coordinates $(\alpha,\beta)$.
The switch from $(x,y)$ to $(\alpha,\beta)$ as independent variables is
allowed wherever the Jacobian of the transformation
$x_{,\alpha}y_{,\beta}-x_{,\beta}y_{,\alpha}\not=0$.

Using (\ref{eq:dydx1}) and (\ref{eq:dydx2}), we obtain two equations
\begin{eqnarray}
({\cal B}+{\cal A}t)y_{,\alpha}-{1\over 2}({\cal C}-2{\cal A}s
-\delta)x_{,\alpha}&=&0
\label{eq:e1}\\
({\cal B}+{\cal A}t)y_{,\beta}-{1\over 2}({\cal C}-2{\cal A}s
+\delta)x_{,\beta}&=&0\,.
\label{eq:e2}
\end{eqnarray}
These two equations together with the identity (\ref{eq:identity})
give us two additional equations
\begin{eqnarray}
{1\over 2}({\cal C}-2{\cal A}s+\delta)y_{,\alpha}-({\cal D}
+{\cal A}r)x_{,\alpha}&=&0
\label{eq:e3}\\
{1\over 2}({\cal C}-2{\cal A}s-\delta)y_{,\beta}-({\cal D}
+{\cal A}r)x_{,\beta}&=&0\,.
\label{eq:e4}
\end{eqnarray}
These are four first-order PDEs in the unknowns $x,y$.
But we must still deal with the unknowns $W,p,q,r,s,t$.

To this end, consider the first- and second-order strip equations.
They are
\begin{eqnarray}
W_{,\alpha}&=&p\,x_{,\alpha}+q\,y_{,\alpha}\\
W_{,\beta}&=&p\,x_{,\beta}+q\,y_{,\beta}
\end{eqnarray}
and
\begin{eqnarray}
p_{,\alpha}&=&r\,x_{,\alpha}+s\,y_{,\alpha}\\
p_{,\beta}&=&r\,x_{,\beta}+s\,y_{,\beta}\\
q_{,\alpha}&=&s\,x_{,\alpha}+t\,y_{,\alpha}\\
q_{,\beta}&=&s\,x_{,\beta}+t\,y_{,\beta}
\end{eqnarray}
respectively.
These equations are needed to guarantee that
\begin{equation}
p=W_{,x}\quad q=W_{,y}\quad r=W_{,xx}\quad s=W_{,xy}\quad t=W_{,yy}\,.
\end{equation}
By substituting the second-order strip equations into
(\ref{eq:e1})-(\ref{eq:e4}) we can eliminate the dependence on $r,s,t$,
and write the results as
\begin{eqnarray}
{\cal B}\,y_{,\alpha}+{\cal A}\,q_{,\alpha}-{1\over 2}({\cal C}
-\delta)\,x_{,\alpha}&=&0\\
{\cal B}\,y_{,\beta}+{\cal A}\,q_{,\beta}-{1\over 2}({\cal C}
+\delta)\,x_{,\beta}&=&0\\
{1\over 2}({\cal C}+\delta)\,y_{,\alpha}-{\cal D}\,x_{,\alpha}
-{\cal A}\,p_{,\alpha}&=&0\\
{1\over 2}({\cal C}-\delta)\,y_{,\beta}-{\cal D}\,x_{,\beta}
-{\cal A}\,p_{,\beta}&=&0\,.
\end{eqnarray}
These are four quasilinear, first-order PDEs in the five unknowns
$(x,y,W,p,q)$.
(Quasilinear in the sense that the equations depend linearly on the
first partial derivatives of $(x,y,W,p,q)$.)
In order to complete the system, we need an additional independent
equation.
We choose this to be the first-order strip equation
\begin{equation}
W_{,\alpha}-p\,x_{,\alpha}-q\,y_{,\alpha}=0\,.
\end{equation}
The above five equations are equations (\ref{eq:cs1})-(\ref{eq:cs5})
in the main text.
This is the desired result.

\subsection*{A.6 Determining $U$ and $V$ via quadratures}

After $W$ has been found as a function of $(x,y)$ we must still determine
the remaining embedding functions $U(x,y),V(x,y)$.
We could in principle solve (\ref{eq:E})-(\ref{eq:G}) directly, but
there is a simpler approach, again due to Darboux, that we use and
outline here.
(For further discussion, see \cite{H&W}.)

Consider a general, positive-definite, 2-dimensional line element
\begin{equation}
ds^2=f_{11}\,dx^2+2f_{12}\,dx\,dy+f_{22}\,dy^2
\label{eq:ds^2f}
\end{equation}
with metric components denoted by $f_{ij}$.
Then the Gaussian curvature $K$ of (\ref{eq:ds^2f}) can be written as
\begin{equation}
\sqrt{f}\,K=(B_{,x}-A_{,y})
\label{eq:KK}
\end{equation}
where $f:=f_{11}f_{22}-(f_{12})^2$ is the determinant of the matrix of
components $f_{ij}$ and
\begin{eqnarray}
A&=&{1\over 2\sqrt{f}}\bigg[\,\bigg({f_{12}\over f_{11}}\bigg)\,f_{11,x}
+f_{11,y}-f_{12,x}\bigg]\label{eq:AA}\\
B&=&{1\over 2\sqrt{f}}\bigg[\,\bigg({f_{12}\over f_{11}}\bigg)\,f_{11,y}
-f_{22,x}+f_{12,y}\bigg]\,.\label{eq:BB}
\end{eqnarray}
Using (\ref{eq:KK}) and Stoke's theorem, we can write
\begin{equation}
\int_\Omega \,\sqrt{f}\,K\,dx\,dy=\oint_{\partial\Omega}(A\,dx+ B\,dy)\,,
\label{eq:intK}
\end{equation}
where the integral on the left hand side of the above equation is
taken over some simply-connected 2-dimensional region $\Omega$.  The
integral on the right hand side is around the closed 1-dimensional
boundary $\partial\Omega$.

If $K=0$ at every point $(x,y)$, the right hand side of (\ref{eq:intK})
vanishes for all closed curves $\partial\Omega$.
This means that there exists a function $\vartheta(x,y)$ such that
\begin{equation}
\vartheta_{,x}=A\quad{\rm and}\quad\vartheta_{,y}=B\,.
\label{eq:thetaxy}
\end{equation}
One can then integrate equations (\ref{eq:thetaxy}) directly to obtain an
expression for $\vartheta$ in terms of the components $f_{ij}$.
The solution is determined up to an overall additive constant.
Moreover, using (\ref{eq:thetaxy}) and the definitions
(\ref{eq:AA}) and (\ref{eq:BB}) of $A$ and $B$, one can show that
\begin{eqnarray}
\bigg(-\sqrt{f_{11}}\sin\vartheta\bigg){}_{,y}
&=&\bigg(-{f_{12}\over\sqrt{f_{11}}}\sin\vartheta
+{\sqrt{f}\over\sqrt{f_{11}}}\cos\vartheta\bigg){}_{,x}\\
\bigg(+\sqrt{f_{11}}\cos\vartheta\bigg){}_{,y}
&=&\bigg(+{f_{12}\over\sqrt{f_{11}}}\cos\vartheta
+{\sqrt{f}\over\sqrt{f_{11}}}\sin\vartheta\bigg){}_{,x}\,.
\end{eqnarray}
This means that there exist functions $U(x,y)$, $V(x,y)$ such that
\begin{eqnarray}
U_{,x}&=&-\sqrt{f_{11}}\sin\vartheta\label{eq:Ux}\\
U_{,y}&=&-{f_{12}\over\sqrt{f_{11}}}\sin\vartheta
+{\sqrt{f}\over\sqrt{f_{11}}}\cos\vartheta\label{eq:Uy}\\
V_{,x}&=&+\sqrt{f_{11}}\cos\vartheta\label{eq:Vx}\\
V_{,y}&=&+{f_{12}\over\sqrt{f_{11}}}\cos\vartheta
+{\sqrt{f}\over\sqrt{f_{11}}}\sin\vartheta\,.\label{eq:Vy}
\end{eqnarray}
These equations can also be integrated directly, allowing us to obtain
expressions for $U,V$ in terms of the components $f_{ij}$.
The solutions for $U,V$ are determined up to overall additive constants.

The freedom in choosing the additive constants for $\vartheta,U,V$
corresponds to the freedom of performing a 2-dimensional Euclidean
motion of the plane.
Such a motion preserves the form of the line element $du^2+dv^2$.
The additive constant for $\vartheta$ corresponds to a rigid rotation of
the plane about the origin.
The additive constants for $U,V$ correspond to a translation.

Given the above results, we now specialize to the case of the flat
2-dimensional line element (\ref{eq:flat2}).
This line element has components
\begin{eqnarray}
f_{11}&=&E-p^2\label{eq:f11}\\
f_{12}&=&F-pq\\
f_{22}&=&G-q^2\label{eq:f22}
\end{eqnarray}
where $p=W_{,x}$ and $q=W_{,y}$.
The solutions for $\vartheta,U,V$ that we find by integrating
(\ref{eq:thetaxy}) and (\ref{eq:Ux})-(\ref{eq:Vy}) (using the above
$f_{ij}$) complete the embedding.

\subsection*{A.7 Gaussian curvature of the Misner 2-geometry}

We start with the Misner 2-geometry
\begin{equation}
ds^{2}=a^{2}\varphi^{4}_{\rm Misner}\left[d\mu^{2}+d\eta^{2}\right]
\label{eq:ds^2app}
\end{equation}
written in terms of the coordinates $(\mu,\eta)$.
To simplify the notation in what follows, we will drop the ``Misner''
subscript from $\varphi_{\rm Misner}$ and define
\begin{equation}
[n]:=\cosh(\mu+2n\mu_0)-\cos\eta.
\end{equation}
Then
\begin{equation}
\varphi:=\varphi_{\rm Misner}=
\sum_{n=-\infty}^\infty\,[n]^{-1/2}\,.
\end{equation}

Our goal is to calculate the Gaussian curvature $K$ of (\ref{eq:ds^2app}),
and to write it in a form which is manifestly everywhere-negative.
If we take $(\mu,\eta)$ as our coordinates $(x,y)$ then
\begin{equation}
E=G=a^2\,\varphi^4\quad{\rm and}\quad F=0
\end{equation}
are the components of the Misner 2-geometry.
For $E=G,\,F=0$, the expression (\ref{eq:Kapp}) for the Gaussian
curvature simplifies considerably:
\begin{equation}
K=-{1\over 2}\,E^{-1}\left[\,(\ln E)_{,xx}+(\ln E)_{,yy}\,\right]\,.
\end{equation}
This is just minus one-half times the covariant Laplacian of $\ln E$.
For $E=a^2\,\varphi^4$, we have
\begin{equation}
K=-2\,a^{-2}\,\varphi^{-4}\left[\,(\ln\varphi)_{,\mu\mu}
+(\ln\varphi)_{,\eta\eta}\,\right]\,.
\end{equation}
This is what we must evaluate.

We begin by calculating the first and second partial derivatives
of $\varphi$.
They are
\begin{eqnarray}
\varphi_{,\mu}&=&\sum_{n=-\infty}^\infty\, -{1\over 2}[n]^{-3/2}
\sinh(\mu+2n\mu_0)\\
\varphi_{,\eta}&=&\sum_{n=-\infty}^\infty\, -{1\over 2}[n]^{-3/2}
\sin\eta\\
\varphi_{,\mu\mu}&=&\sum_{n=-\infty}^\infty\,\bigg(-{1\over 2}[n]^{-3/2}
\cosh(\mu+2n\mu_0)\nonumber\\
& &\quad\quad\quad\quad\quad\quad\quad
+{3\over 4}[n]^{-5/2}\sinh^2(\mu+2n\mu_0)\,\bigg)\\
\varphi_{,\eta\eta}&=&\sum_{n=-\infty}^\infty\,\bigg(-{1\over 2}[n]^{-3/2}
\cos\eta+{3\over 4}[n]^{-5/2}\sin^2\eta\,\bigg)\,.
\end{eqnarray}
Since
\begin{equation}
(\ln \varphi)_{,\mu\mu}+(\ln \varphi)_{,\eta\eta}=\varphi^{-2}\,
\left[\,\varphi\,\varphi_{,\mu\mu}+\varphi\,\varphi_{,\eta\eta}
-(\varphi_{,\mu})^2-(\varphi_{,\eta})^2\,\right]\,,
\label{eq:log}
\end{equation}
it is also convenient to evaluate
\begin{eqnarray}
\lefteqn{\varphi\,\varphi_{,\mu\mu}+\varphi\,\varphi_{,\eta\eta}}
\nonumber\\
&&={1\over 4}\sum_{m=-\infty}^\infty\sum_{n=-\infty}^\infty\,
{\cosh(\mu+2m\mu_0)\cosh(\mu+2n\mu_0)-\cos^2\eta\over
[m]^{3/2}\,[n]^{3/2}}
\label{eq:part1}\\
\lefteqn{(\varphi_{,\mu})^2+(\varphi_{,\eta})^2}\nonumber\\
&&={1\over 4}\sum_{m=-\infty}^\infty\sum_{n=-\infty}^\infty\,
{\sinh(\mu+2m\mu_0)\sinh(\mu+2n\mu_0)+\sin^2\eta\over
[m]^{3/2}\,[n]^{3/2}}\,.
\label{eq:part2}
\end{eqnarray}
Combining (\ref{eq:log}), (\ref{eq:part1}), and (\ref{eq:part2}) we see
that
\begin{eqnarray}
\lefteqn{(\ln\varphi)_{,\mu\mu}+(\ln\varphi)_{,\eta\eta}}\nonumber\\
&&=\varphi^{-2}\,\left[\,\varphi\,\varphi_{,\mu\mu}
+\varphi\,\varphi_{,\eta\eta}
-(\varphi_{,\mu})^2-(\varphi_{,\eta})^2\,\right]\nonumber\\
&&={1\over 4}\,\varphi^{-2}
\sum_{m=-\infty}^\infty\sum_{n=-\infty}^\infty\,
{\cosh(2\mu_0(m-n))-1\over [m]^{3/2}\,[n]^{3/2}}\,.
\label{eq:temp}
\end{eqnarray}
{}From this we get the final result
\begin{equation}
K=-{1\over 2}\,a^{-2}\,\varphi^{-6}
\sum_{m=-\infty}^\infty\sum_{n=-\infty}^\infty
{\cosh\left(2\mu_0(m-n)\right)-1\over[m]^{3/2}\,[n]^{3/2}}\,.
\end{equation}
This is the manisfestly negative-definite expression for the Gaussian
curvature given in the main text.

One may rightfully worry about those places where
\begin{equation}
[n]:=\cosh(\mu+2n\mu_0)-\cos\eta=0\,.
\end{equation}
When this happens, $\varphi=\sum\,[n]^{-1/2}\rightarrow\infty$ and
$K\rightarrow 0$.
This occurs only if
\begin{equation}
\eta=0\quad{\rm and}\quad\mu+2n\mu_0=0
\end{equation}
where $n$ is an integer.
For $-\mu_0\leq\mu\leq\mu_0$, we see that the second equality holds
only if $\mu=0$ and $n=0$.
Thus, $[n]=0$ if and only if $(\mu,\eta)=(0,0)$.
This is the ``point'' at spatial infinity, and there we expect the
2-geometry corresponding to the Misner data to be flat.

\newpage
\section*{Figure Captions}

\noindent \vspace{7pt}
Fig.~1: \begin{minipage}[t]{5in}
The net of characteristics propagating inward from the $R=const$ initial
value surface for $\mu_0=1,2,3$.
\end{minipage}

\vspace{7pt}
\noindent \vspace{7pt}
Fig.~2: \begin{minipage}[t]{5in}
The principal directions of the extrinsic curvature and the characteristic
directions near the center.
\end{minipage}

\vspace{7pt}
\noindent \vspace{7pt}
Fig.~3: \begin{minipage}[t]{5in}
The computed crossing of two characteristics of the same family for
$\mu_0=6$.
\end{minipage}

\vspace{7pt}
\noindent \vspace{7pt}
Fig.~4: \begin{minipage}[t]{5in}
Shapes of 3-dimensional embedding diagrams when viewed along the $v$ axis
for $\mu_0=1,3$.
\end{minipage}

\vspace{7pt}
\noindent \vspace{7pt}
Fig.~5: \begin{minipage}[t]{5in}
Perspective view of the (incomplete) embedding diagram for $\mu_0=3$.
\end{minipage}

\vspace{7pt}
\noindent \vspace{7pt}
Fig.~6: \begin{minipage}[t]{5in}
Constant $W$ contours of the embedding diagrams for $\mu_0=1.5,2,2.5$.
\end{minipage}


\newpage
\begin{thebibliography}{99}

\bibitem{annetal}
Anninos P, Hobill D, Seidel E, Smarr L, and Suen W-M
1993
``The Collision of Two Black Holes''
{\it Phys. Rev. Lett.} {\bf 71} 2851-2854

\bibitem{PP}
Price R H and Pullin J
1994
``Colliding Black Holes: The Close Limit''
{\it Phys. Rev. Lett.} {\bf 72} 3297-3300;
Abrahams A M and Cook G B
1994
``Collisions of Boosted Black Holes: Perturbation Theory Prediction
of Gravaitational Radiation''
{\it Phys. Rev.} {\bf D50} R2364-2367

\bibitem{Misner}
Misner C W
1960
``Wormhole Initial Conditions''
{\it Phys. Rev.} {\bf 118} 1110-1111

\bibitem{Spivak5}
Spivak M
1975
{\it A Comprehensive Introduction to Differential Geometry,
Volume 5}
(Boston: Publish or Perish) Chap 11

\bibitem{C&H2}
Courant R and Hilbert D
1962
{\it Methods of Mathematical Physics,
Volume II: Partial Differential Equations}
(New York: Interscience) Appendix 1 to Chap 5

\bibitem{H&W}
Hartman P and Wintner A
1951
``Gaussian Curvature and Local Embedding''
{\it Am. Jour. of Math.} {\bf 73} 876-884

\bibitem{P&S}
Poznyak \'{E} G and Shikin E V
1976
``Surfaces of Negative Curvature''
{\it Jour. of Soviet Math.} {\bf 5} 865-887

\bibitem{Bernstein}
Bernstein D
1994
``An Iterative Scheme for Computing Isometric Embedding Diagrams for
use in Numerical Relativity Calculations''
University of New England, Australia Preprint

\bibitem{Hsiung}
Hsiung C-C
1981
{\it A First Course in Differential Geometry}
(New York: Wiley) pp 264-267

\bibitem{whitham}
Whitham G B
1974
{\it Linear and Nonlinear Waves}
(New York: Wiley) Chap 2

\bibitem{sunprice2}
Sun Y and Price R H
1990
``Excitation of Schwarzschild Quasinormal Modes by Collapse''
{\it Phys. Rev.} {\bf D41} 2492-2506




\end{thebibliography}
\end{document}